\documentclass[conference]{IEEEtran}
\IEEEoverridecommandlockouts

\usepackage{cite}
\usepackage{amsmath,amssymb,amsfonts}
\usepackage{algorithmic}
\usepackage{graphicx}
\usepackage{textcomp}
\usepackage{xcolor}
\def\BibTeX{{\rm B\kern-.05em{\sc i\kern-.025em b}\kern-.08em
    T\kern-.1667em\lower.7ex\hbox{E}\kern-.125emX}}

\usepackage{algorithm}
\usepackage{array}
\usepackage{balance}
\usepackage{stfloats}
\usepackage{url}
\usepackage{verbatim}
\usepackage{cite}
\usepackage[most]{tcolorbox}
\usepackage{listings}
\usepackage{threeparttable}
\usepackage{graphicx}
\usepackage{subfigure}
\usepackage{booktabs}
\usepackage{multirow}

\usepackage[table,dvipsnames]{xcolor}
\usepackage{array}

\begin{document}


\title{DMFI: A Dual-Modality Log Analysis Framework for Insider Threat Detection with LoRA-Tuned Language Models}



\author{
\IEEEauthorblockN{
Kaichuan Kong\IEEEauthorrefmark{1},
Dongjie Liu\IEEEauthorrefmark{1},
Xiaobo Jin\IEEEauthorrefmark{2}\thanks{Corresponding author: xiaobo.jin@xjtlu.edu.cn},
Guanggang Geng\IEEEauthorrefmark{1},
Zhiying Li\IEEEauthorrefmark{1},
Jian Weng\IEEEauthorrefmark{1}
}
\IEEEauthorblockA{\IEEEauthorrefmark{1}College of Cyber Security, Jinan University, Guangzhou, China}
\IEEEauthorblockA{\IEEEauthorrefmark{2}School of Advanced Technology, Xi’an Jiaotong-Liverpool University, Suzhou, China}
\IEEEauthorblockA{Emails: willkkc@stu2021.jnu.edu.cn, tzezd2019@stu2020.jnu.edu.cn, \\
\{djliu, gggeng\}@jnu.edu.cn, xiaobo.jin@xjtlu.edu.cn, cryptjweng@gmail.com}}


\maketitle

\begin{abstract}
Insider threat detection (ITD) poses a persistent and high-impact challenge in cybersecurity due to the subtle, long-term, and context-dependent nature of malicious insider behaviors. Traditional models often struggle to capture semantic intent and complex behavior dynamics, while existing LLM-based solutions face limitations in prompt adaptability and modality coverage.
To bridge this gap, we propose DMFI, a dual-modality framework that integrates semantic inference with behavior-aware fine-tuning. DMFI converts raw logs into two structured views: (1) a semantic view that processes content-rich artifacts (e.g., emails, https) using instruction-formatted prompts; and (2) a behavioral abstraction, constructed via a 4W-guided (When-Where-What-Which) transformation to encode contextual action sequences. Two LoRA-enhanced LLMs are fine-tuned independently, and their outputs are fused via a lightweight MLP-based decision module.
We further introduce DMFI-B, a discriminative adaptation strategy that separates normal and abnormal behavior representations, improving robustness under severe class imbalance.
Experiments on CERT r4.2 and r5.2 datasets demonstrate that DMFI outperforms state-of-the-art methods in detection accuracy.  Our approach combines the semantic reasoning power of LLMs with structured behavior modeling, offering a scalable and effective solution for real-world insider threat detection.

\end{abstract}


\section{Introduction}
\label{Introduction}

Insider threats pose a severe and persistent challenge to cybersecurity due to their inherent stealth and legitimacy. Unlike external adversaries, insiders often possess authorized credentials and operational familiarity, allowing them to operate within acceptable behavioral bounds while evading traditional security mechanisms~\cite{fei2025laaeb}. The consequences of insider threats range from intellectual property theft to critical infrastructure sabotage, often resulting in long-term reputational and financial damage. Despite the widespread adoption of rule-based systems, access controls, and behavioral monitoring techniques, real-time and accurate detection of insider threats remains a challenge, primarily due to the adaptive, multimodal, and context-dependent nature of human behavior~\cite{inayat2024insider}.

Research in insider threat detection (ITD) has evolved from static rule-based heuristics to statistical models and deep learning (DL) approaches~\cite{liu2018detecting, le2021anomaly, al2020review}. Traditional supervised methods, such as SVM~\cite{le2020analyzing} and XGBoost~\cite{kan2023data}, along with unsupervised models like Isolation Forest~\cite{le2021anomaly}, have demonstrated utility in analyzing structured audit data. Deep neural networks, including LSTM, CNN~\cite{al2024comparative}, and GNN~\cite{xiao2024sentinel}, further enhance modeling of temporal and relational dependencies. However, rule-based systems rely on static policies and lack adaptability to evolving attack patterns. Similarly, conventional machine learning methods typically treat behaviors as numerical features, overlooking the semantic context of actions. This abstraction limits their reasoning capability and generalization to novel or context-rich insider strategies~\cite{yuan2021deep}.

Recent advances in large language models (LLMs) have introduced a transformative shift in ITD research, driven by LLMs' strengths in semantic understanding, contextual reasoning, and natural language generation~\cite{chen2024survey}. Existing LLM-based solutions typically fall into two categories: prompt-based and fine-tuning-based methods. Prompt-based systems, such as LogGPT~\cite{qi2023loggpt}, RedChronos~\cite{li2025redchronos}, and Audit-LLM~\cite{song2024audit}, apply pretrained LLMs to log sequences using handcrafted prompts. These models offer high interpretability and require no retraining, making them particularly attractive for low-resource or privacy-sensitive environments. However, their detection performance is often sensitive to prompt design and lacks behavioral adaptability.

Fine-tuning-based approaches, such as Confront-LLM~\cite{song2025confront}, LLM4Sec~\cite{karlsen2024benchmarking}, and FedITD~\cite{wang2024feditd}, are designed to adapt LLMs to structured behavioral datasets for more accurate modeling of user intent and anomaly patterns. These methods often offer superior performance and more granular threat detection capabilities. However, they typically require extensive supervision and tend to overlook the semantic richness embedded in natural language communication logs. Moreover, most prior work treats either semantic artifacts (e.g., emails, HTTP requests) or behavioral sequences (e.g., login-event graphs) in isolation, without integrating both modalities in a coherent framework.

Despite these advances, a fundamental gap persists: existing methods rarely capture both the \textbf{semantic signals} present in communication logs and the \textbf{temporal structure} of user behavior in a cohesive manner, especially under class imbalance and limited supervision. Prompt-based LLMs tend to lack dynamic behavioral awareness, while fine-tuning methods frequently discard textual richness. Additionally, many existing approaches adopt monolithic modeling strategies that conflate heterogeneous inputs or employ coarse-grained scoring mechanisms, which constrain both interpretability and detection granularity.

To address these limitations, we propose \textbf{DMFI} (\textit{Dual-Modality Fine-tuned Inference})\textemdash a unified framework that integrates semantic reasoning and behavioral abstraction using instruction-tuned LLMs. DMFI introduces several innovations:

First, we introduce a \textbf{4W-guided behavior abstraction} that transforms structured logs into concise, interpretable natural language summaries aligned with "When–Where-What–Which" dimensions. This design compresses redundant activity records while enhancing temporal reasoning. Second, we construct \textbf{modality-specific prompts} and apply LoRA-based parameter-efficient fine-tuning to adapt two LLM branches (semantic and behavioral) under low-resource constraints. Third, we develop a \textbf{discriminative dual-branch strategy} (DMFI-B) that separately fine-tunes on normal and abnormal data, enabling margin-based risk estimation with heightened sensitivity to rare threats. Finally, a lightweight decision module fuses statistical semantic scores with behavioral LLM outputs, yielding calibrated predictions and interpretable justifications.

\textbf{Our contributions are summarized as follows:}
\begin{itemize}
\item We propose \textbf{DMFI}, a dual-modality insider threat detection framework that unifies prompt-driven semantic analysis and fine-tuned behavioral profiling using modality-aligned LLMs.
\item We introduce a novel \textbf{4W-guided abstraction and multi-statistical scoring} mechanism to represent heterogeneous behavioral and semantic cues in a unified format, enabling session-level reasoning across variable-length and multi-modal data.
\item We explore two LoRA-based fine-tuning paradigms, including a \textbf{discriminative dual-branch strategy} (DMFI-B), which enhances anomaly separability via contrastive scoring between normal and abnormal models.
\item Extensive experiments on CERT r4.2 and r5.2 demonstrate that DMFI outperforms both traditional and LLM-based baselines in terms of precision, robustness, and interpretability, while remaining resource-efficient and deployment-friendly.
\end{itemize}

The remainder of this paper is structured as follows: 
Section~\ref{section:related_work} reviews conventional insider threat detection techniques and recent advances in LLM-based paradigms, highlighting the gaps our work addresses. 
Section~\ref{section:proposed_architecture} presents the overall architecture of our proposed DMFI framework, including the preprocessing pipeline, prompt construction, and fine-tuning strategies.
Section~\ref{section:experiments} reports our experimental settings, baseline comparisons, ablation studies, and efficiency-aware analysis.
Finally, Section~\ref{section:conclusion} concludes the paper and outlines directions for future work.

\begin{figure*}[!t]
\centering
\includegraphics[width=0.92\linewidth]{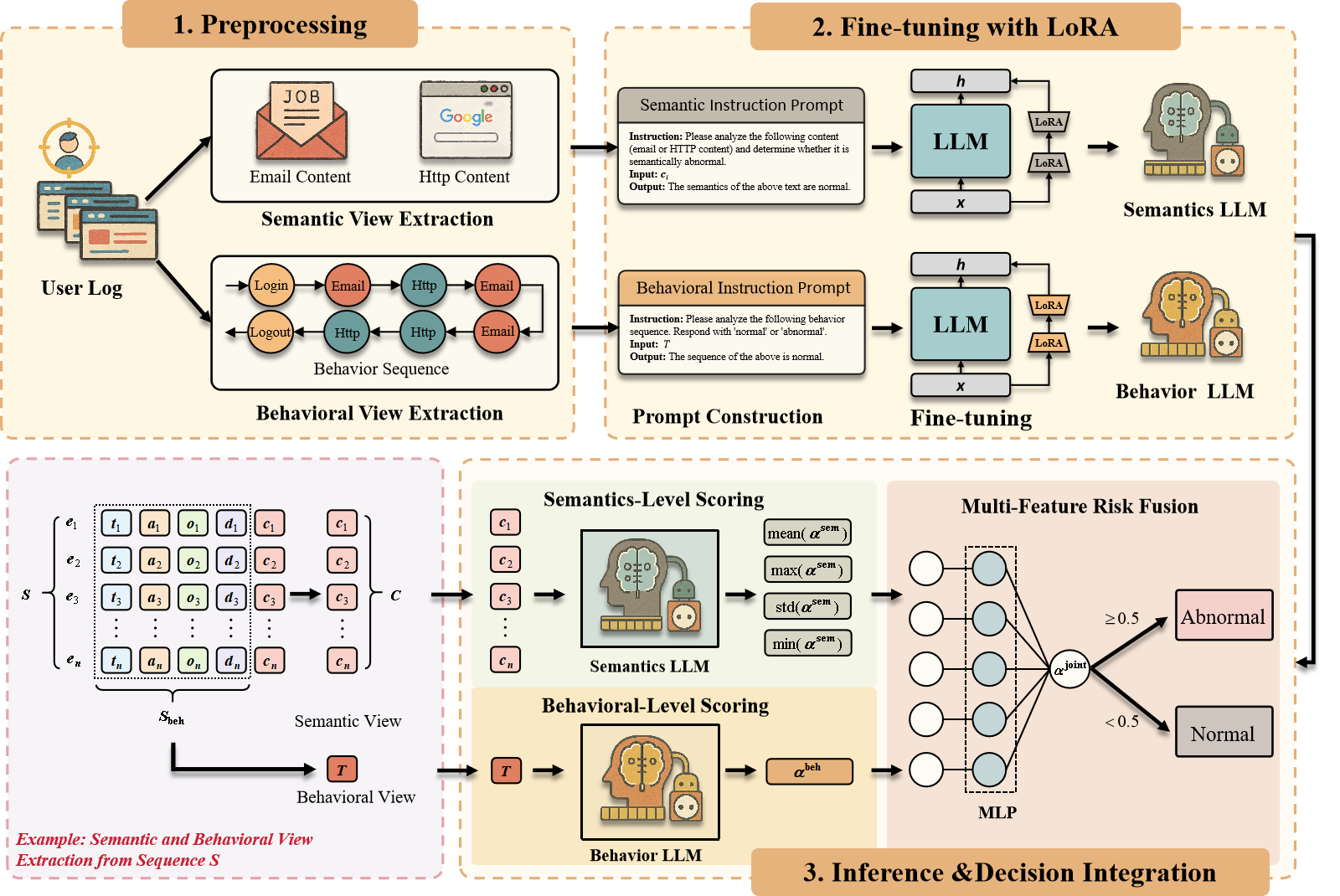}
\caption{
Overview of DMFI. 
It consists of three main components: 
(1) \textbf{Preprocessing}, which extracts semantic and behavioral views from user logs;
(2) \textbf{Fine-tuning with LoRA}, where dual instruction-tuned LLMs are trained on semantic and behavioral prompts; and 
(3) \textbf{Inference and Decision Integration}, where multi-feature scores are fused via a lightweight MLP to derive a final anomaly score $\alpha_{\text{joint}}$ for insider threat detection.
}
\label{fig:architecture}
\vspace{-1em}
\end{figure*}

\section{Related Work}
\label{section:related_work}

We categorize prior research on insider threat detection (ITD) into three major paradigms: (1) traditional rule- or model-based detection, (2) prompt-based inference with large language models (LLMs), and (3) fine-tuning-based LLM adaptation. We summarize representative approaches and highlight key limitations to motivate our proposed dual-modality framework.

\subsection{Traditional Methods for Insider Threat Detection}

Early ITD techniques span rule-based policies and statistical profiling. Signature- or rule-based systems~\cite{homoliak2019insight} depend on predefined heuristics and thresholds, which offer high interpretability but are fragile in the face of mimicry and low-rate anomalies. Statistical profiling~\cite{liu2018detecting} captures user baselines but struggles with stealthy, context-dependent deviations.

Machine learning (ML) methods provide automatic pattern recognition but often depend on hand-crafted features. Classical classifiers such as SVM~\cite{le2020analyzing} and XGBoost~\cite{kan2023data} offer fast inference but lack temporal or semantic context. Unsupervised models like Isolation Forest~\cite{le2021anomaly} generalize better but frequently mislabel outliers.

Recent deep learning (DL) methods improve representation learning. RNN-based models (e.g., LSTM~\cite{villarreal2021hunting}) capture sequential patterns, while CNNs~\cite{al2024comparative} and autoencoders~\cite{zhu2024auth} learn latent structures. Graph-based approaches such as Sentinel~\cite{xiao2024sentinel} and DeepAudit~\cite{yuan2021deep} model interactions across users and systems but still treat actions as symbolic units without semantic abstraction.

\textit{Limitation:} Traditional approaches often lack the capacity to reason over semantic content or model complex behavioral intent—both critical for identifying covert insider threats.

\subsection{LLM-Based Detection Paradigms}

The emergence of LLMs has enabled semantic understanding and reasoning over unstructured logs. Two mainstream paradigms have emerged: prompt-based inference and parameter-efficient fine-tuning.

\subsubsection{Prompt-Based LLMs}

Prompt-based methods utilize pretrained LLMs in a zero-/few-shot setting without task-specific training. LogGPT~\cite{qi2023loggpt} reformats logs as structured prompts for GPT-style models, while RedChronos~\cite{li2025redchronos} employs prompt ensembles and evolutionary strategies to improve robustness. Audit-LLM~\cite{song2024audit} introduces multi-agent CoT-style debates to improve consistency. Other studies~\cite{clairoux2024use, gelman2025scalable} explore prompts for insider sentiment analysis or cyber threat intelligence extraction.

\textit{Limitation:} Despite flexibility, prompt-based methods are highly sensitive to template design, offer limited behavior abstraction, and lack stability under evolving log semantics or class imbalance.

\subsubsection{Fine-Tuned LLMs}

Fine-tuning adapts LLMs to domain-specific data, allowing deeper alignment with task semantics and temporal dependencies. Confront-LLM~\cite{song2025confront} encodes audit logs as narratives and applies contrastive learning to improve anomaly discrimination. Karlsen et al.~\cite{karlsen2024benchmarking} fine-tune DistilRoBERTa on log datasets and explore internal decision structures using SHAP and t-SNE. FedITD~\cite{wang2024feditd} integrates LoRA with federated optimization to support decentralized, privacy-aware adaptation.

\textit{Limitation:} Most fine-tuning approaches are unimodal—focusing either on content semantics or behavioral patterns—without a unified framework that fuses both. Additionally, interpretability and output controllability remain limited.

\subsection{Positioning of This Work}

To address the limitations of prior methods, we propose a dual-modality fine-tuning framework that jointly models semantic content and behavioral abstraction. By applying instruction-style supervised fine-tuning with modality-specific LoRA adapters, our approach supports interpretable, context-aware, and parameter-efficient threat detection.
Compared with prompt-based methods, our framework improves robustness under user variability and class imbalance. It further outperforms unimodal fine-tuning by integrating cross-modal signals for calibrated, analyst-aligned risk estimation, offering a unified and deployable solution for insider threat detection.

\section{Proposed Architecture}
\label{section:proposed_architecture}

\subsection{Overview}

To enable accurate and interpretable insider threat detection, we propose \textbf{DMFI}, a dual-perspective detection architecture that combines semantic understanding and behavioral profiling using large language models (LLMs). As shown in Fig.~\ref{fig:architecture}, the framework comprises three stages: \textit{modality-aware preprocessing}, \textit{modality-specific modeling}, and \textit{multi-source inference integration}.

In the preprocessing stage, raw audit logs are parsed into structured event tuples and decomposed into two complementary views: a \textit{semantic view}, which captures rich textual content such as emails and web documents, and a \textit{behavioral view}, which summarizes user activity using a lightweight 4W-guided abstraction. Each view is converted into instruction-style prompts suitable for LLM-based evaluation.
Next, we employ parameter-efficient fine-tuning using Low-Rank Adaptation (LoRA), applied independently to two instruction-style LLMs—one per modality. We explore two fine-tuning strategies: DMFI-A (unified) and DMFI-B (dual-branch). DMFI-A directly fine-tunes a single model on mixed data, while DMFI-B performs discriminative adaptation using separate models for normal and abnormal data, enabling more refined anomaly sensitivity under imbalance conditions.
Finally, at inference time, semantic scores from multiple log entries are aggregated via descriptive statistics (mean, max, std, min) to produce a fixed-size semantic representation. This, combined with the behavioral-level risk score, forms a multi-feature vector which is passed through a lightweight MLP for final decision making. The output includes a calibrated anomaly score and binary label, along with an optional natural language explanation to aid analyst interpretation.

Overall, DMFI combines structured behavior modeling and semantic reasoning in a scalable and modular framework, supporting fine-grained, multimodal threat detection across diverse enterprise environments.

\subsection{Preprocessing}
\label{sec:preprocessing}

\begin{figure*}[!t]
\centering
\includegraphics[width=0.75\linewidth]{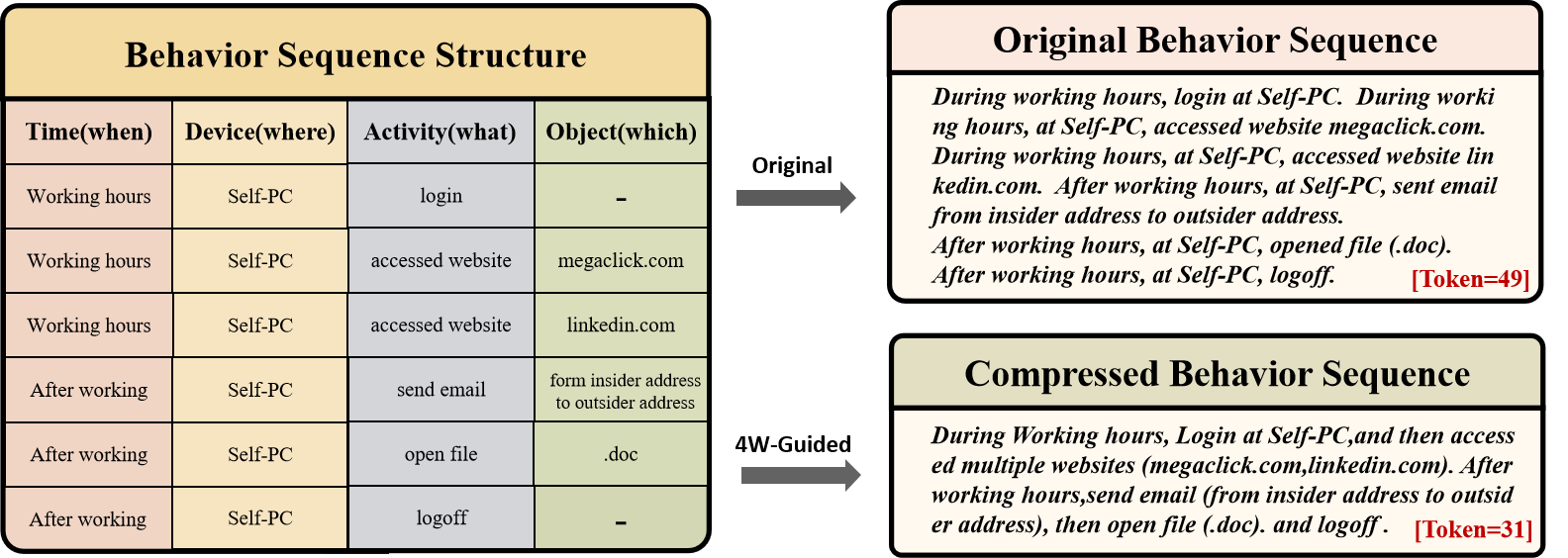}
\caption{
Illustrative example of behavior sequence compression. The left table organizes raw user actions using the 4W schema (\textit{When}, \textit{Where}, \textit{What}, \textit{Which}).
On the right, we compare the original verbose sequence with a compressed version generated by our 4W-guided abstraction strategy.
While the original form enumerates each atomic behavior separately, the compressed version merges related actions into concise natural language.
This reduces token length (from 49 to 31 in this case) while retaining key behavioral semantics.
}
\label{fig:compress}
\vspace{-1em}
\end{figure*}

To support dual-perspective modeling in insider threat detection, we design a structured preprocessing pipeline that transforms raw enterprise audit logs into two modality-specific representations: a \textit{semantic view} and a \textit{behavioral view}. These representations serve as the foundational inputs for downstream prompt construction and LLM reasoning.

Each log record is parsed into a structured tuple:
\begin{equation}
e_i = (t_i, a_i, o_i, d_i, c_i),
\end{equation}
where $t_i$ is the timestamp (e.g., work hours), $a_i$ denotes the action type (e.g., login, send\_email), $o_i$ indicates the target object (e.g., file name, domain), $d_i$ specifies the device, and $c_i$ captures optional semantic content (e.g., email body, URL text). The full log sequence $S = \{e_1, e_2, ..., e_n\}$ is decomposed into two parallel views, as described below.

\subsubsection{Semantic View: Log Content Extraction}
\label{sec:semantic-view}

\begin{algorithm}[!t]
\small
\caption{4W-Guided Behavior Abstraction}
\label{alg:semanticmapping}
\begin{algorithmic}[1]
\REQUIRE Structured sequence $\mathcal{S}_{\text{beh}} = \{e_1^{\text{beh}}, e_2^{\text{beh}}, \ldots, e_n^{\text{beh}}\}$
\ENSURE Interpretable natural language sequence $T$

\STATE $T \leftarrow \emptyset$ \COMMENT{Initialize the output sentence list}
\STATE $G_{\text{time}} \leftarrow$ \textbf{GroupByTime}($\mathcal{S}_{\text{beh}}$) \COMMENT{Segment sequence by temporal context}

\FOR{each time-slice $S_t$ in $G_{\text{time}}$}
    \STATE $G_{\text{device}} \leftarrow$ \textbf{GroupByDevice}($S_t$) \COMMENT{Further split by device within each time group}
    
    \FOR{each device-group $S_{t,d}$ in $G_{\text{device}}$}
        \STATE $w \leftarrow$ \textbf{ExtractTime}($S_{t,d}$) \COMMENT{\textbf{When}: representative time (e.g., "after hours")}
        \STATE $l \leftarrow$ \textbf{ExtractDevice}($S_{t,d}$) \COMMENT{\textbf{Where}: device or platform}
        \STATE $\mathbb{A} \leftarrow$ \textbf{ExtractActions}($S_{t,d}$) \COMMENT{\textbf{What}: user actions in group}
        \STATE $\mathbb{O} \leftarrow$ \textbf{ExtractObjects}($S_{t,d}$) \COMMENT{\textbf{Which}: operated or accessed objects}
        
        \STATE $s \leftarrow$ \textbf{Render4WSentence}($w, l, \mathbb{A}, \mathbb{O}$) \COMMENT{Construct sentence via 4W template}
        \STATE Append $s$ to $T$
    \ENDFOR
\ENDFOR

\RETURN $T$
\end{algorithmic}
\end{algorithm}

The semantic view focuses on extracting meaningful natural language artifacts embedded within system activities, such as email messages and web access logs. We filter out all log entries with empty content fields to construct the semantic sequence:
\begin{equation}
\mathcal{C} = \{c_i \mid e_i = (t_i, a_i, o_i, d_i, c_i) \in S,\ c_i \neq \emptyset\},
\end{equation}
Each $c_i$ is later transformed into an instruction-style prompt to support content-level anomaly detection tasks such as intent deviation, sensitive data exposure, or phishing behavior, handled by the semantic branch of our framework.

\subsubsection{Behavioral View: 4W-Guided Abstraction}
\label{sec:behavioral-view}

The behavioral view captures structural user activity patterns by discarding the semantic content $c_i$ and retaining operational fields:
\begin{equation}
e_i^{\text{beh}} = (t_i, a_i, o_i, d_i), \quad
\mathcal{S}_{\text{beh}} = \{e_1^{\text{beh}}, ..., e_n^{\text{beh}}\}.
\end{equation}


To reduce input redundancy and emphasize salient behavior, we propose a lightweight \textbf{4W-guided abstraction} inspired by the classic 5W1H framework~\cite{narvala2023identifying}. Our abstraction focuses on four directly observable dimensions from structured logs—\textit{When}, \textit{Where}, \textit{What}, and \textit{Which}—motivated by the following considerations:

\begin{itemize}
    \item \textbf{Observability:} \textit{When}, \textit{Where}, and \textit{What} correspond to log fields such as timestamps, device IDs, and action types.
    
    \item \textbf{Object-centricity:} \textit{Which} captures the manipulated object (e.g., file, URL, or email), offering finer-grained behavioral semantics and replacing the implicit \textit{Who} in user-level aggregation.
    
    \item \textbf{Abstraction scope:} We exclude \textit{Who}, \textit{Why}, and \textit{How} due to their reliance on external context or intent-level reasoning, which is beyond the scope of structured behavior modeling.
    
    \item \textbf{Efficiency:} This abstraction balances expressiveness and compactness, preserving anomalous cues while reducing redundancy and noise.
\end{itemize}
 
Routine actions (e.g., browsing) are compressed, while abnormal patterns (e.g., off-hour access or device switching) are retained in detail. The abstraction function:
\begin{equation}
T = \mathcal{F}_{\text{4W}}(\mathcal{S}_{\text{beh}}),
\end{equation}
translates $\mathcal{S}_{\text{beh}}$ into a structured sequence $T$ of narrative sentences, grouped by time and context. 
The resulting $T$ sequence is used as input to the behavioral instruction-tuned LLM. Algorithm~\ref{alg:semanticmapping} details the sentence rendering procedure, while Fig.~\ref{fig:compress} provides a concrete example using real log data.

\subsection{Fine-tuning with LoRA}
To adapt large language models (LLMs) for insider threat detection across semantic and behavioral modalities, we employ supervised fine-tuning (SFT) enhanced with Low-Rank Adaptation (LoRA)~\cite{mao2025survey}. This approach enables efficient domain adaptation of instruction-following LLMs using structured audit logs while significantly reducing parameter overhead.
In our framework, two modality-specific LLMs are independently fine-tuned based on structured input-output pairs derived from the preprocessing pipeline. The task is formulated as a binary classification problem, where each input is labeled as either \textit{Normal} or \textit{Abnormal} according to dataset annotations.

\subsubsection{Modality-Specific Prompt Construction}

We design instruction-style prompts that align with the behavior of instruction-tuned LLMs, mapping raw log content or behavior sequences to a judgment task.

\paragraph{Semantic Prompting}

For each semantic unit $c_i \in \mathcal{C}$ (e.g., email body, HTTP request), we construct an instruction-aligned prompt as follows:

\begin{quote}
\small
\textit{\textbf{Instruction:} Please analyze the following content (email or HTTP message) and determine whether it is semantically abnormal. Respond with both an anomaly score and a classification result.} \\
\textit{\textbf{Input:} $c_i$} \\
\textit{\textbf{Output:} Anomaly Score = ..., Prediction = ``Normal'' / ``Abnormal''.}
\end{quote}

This guides the semantic LLM to learn linguistic anomaly patterns such as intent inconsistency, phishing content, or content-sensitive violations. 

\paragraph{Behavioral Prompting}

Each behavior sequence $T$, abstracted via the 4W rendering function, is treated as a temporal session narrative. The corresponding prompt is formulated as:

\begin{quote}
\small
\textit{\textbf{Instruction:} Please analyze the following behavior sequence. Respond with both an anomaly score and a classification result ("Normal" or "Abnormal").} \\
\textit{\textbf{Input:} $T$} \\
\textit{\textbf{Output:} Anomaly Score = ..., Prediction = ``Normal'' / ``Abnormal''.}
\end{quote}

This enables the behavioral LLM to model high-level user behavior chains and detect anomalous deviations across sessions.

\subsubsection{Discriminative Fine-Tuning Strategies with LoRA}

Each LLM is initialized from a pre-trained checkpoint and adapted to the insider threat detection task using Low-Rank Adaptation (LoRA). Specifically, low-rank matrices $\Delta W$ are injected into the attention and feed-forward layers of the base model to enable parameter-efficient fine-tuning under limited data conditions.

Let $\mathcal{D} = \{(x_i, y_i)\}$ denote the supervised training dataset, where $x_i$ is the formatted input and $y_i \in \{0,1\}$ indicates whether the instance is normal or abnormal. For the convenience of discussion, we define $\mathcal{D}_{\text{norm}} = \{(x_i, y_i)|y_i = 0\}$ and $\mathcal{D}_{\text{abn}} = \{(x_i, y_i)|y_i = 1\}$.

We explore two fine-tuning strategies—DMFI-A and DMFI-B—that influence the model structure and scoring formulation.

\paragraph{DMFI-A: Unified Fine-Tuning}
DMFI-A employs a unified training strategy, in which both normal and abnormal sessions are jointly used to fine-tune a single LoRA-augmented model. The combined training dataset is defined as:
\begin{equation}
\mathcal{D}_{\text{mix}} = \mathcal{D}_{\text{norm}} \cup \mathcal{D}_{\text{abn}}.
\end{equation}

The unified model is fine-tuned end-to-end:
\begin{equation}
\mathcal{M}_{\text{mix}} = \texttt{FineTune}(\mathcal{M}_0, \mathcal{D}_{\text{mix}};\, \Delta W_{\text{mix}}).
\end{equation}

At inference time, a direct anomaly confidence score is computed as:
\begin{equation}
\alpha = \mathcal{M}_{\text{mix}}(x).
\end{equation}

Here, $\alpha \in [0,1]$ indicates the model's estimated probability of the input $x$ being abnormal.

\paragraph{DMFI-B: Discriminative Dual-Branch Fine-Tuning}
To explicitly disentangle normal and abnormal representations, DMFI-B fine-tunes two independent models from the same pretrained backbone $\mathcal{M}_0$:
\begin{align}
\mathcal{M}_{\text{norm}} &= \texttt{FineTune}(\mathcal{M}_0, \mathcal{D}_{\text{norm}};\, \Delta W_{\text{norm}}), \\
\mathcal{M}_{\text{abn}}  &= \texttt{FineTune}(\mathcal{M}_0, \mathcal{D}_{\text{abn}};\, \Delta W_{\text{abn}}),
\end{align}
where both models are used to evaluate the same input $x$, each producing a confidence score.

The final anomaly probability is calculated using a margin-based formulation:
\begin{equation}
\alpha = \sigma\left( \mathcal{M}_{\text{abn}}(x) - \mathcal{M}_{\text{norm}}(x) \right),
\end{equation}
where $\sigma(\cdot)$ denotes the sigmoid function. This formulation leverages the relative discrepancy between abnormal and normal model responses to highlight anomaly-sensitive signals in a normalized manner.

\begin{algorithm}[!t]
\small
\caption{Multi-Modality Threat Inference}
\label{alg:inference}
\begin{algorithmic}[1]
\REQUIRE Event sequence $S = \{e_i = (t_i, a_i, o_i, d_i, c_i)\}_{i=1}^n$
\ENSURE Threat label $y^{\text{final}}$

\STATE $\mathcal{C} \leftarrow \{c_i \mid e_i = (t_i, a_i, o_i, d_i, c_i) \in S,\ c_i \neq \emptyset\}$
\STATE $T \leftarrow \mathcal{F}_{\text{4W}}(S)$

\FOR{each $c_i \in \mathcal{C}$}
    \STATE $\alpha^{\text{sem}}_i \gets f_{\text{sem}}(c_i)$
\ENDFOR

\STATE $\alpha^{\text{sem}} \gets \{\alpha^{\text{sem}}_i \}_{i=1}^{|\mathcal{C}|}$
\STATE $\mathbf{v}^{\text{sem}} \gets \left[
    \texttt{mean}(\alpha^{\text{sem}}),
    \texttt{max}(\alpha^{\text{sem}}), 
    \texttt{std}(\alpha^{\text{sem}}),
    \texttt{min}(\alpha^{\text{sem}})
\right]$

\STATE $\alpha^{\text{beh}} \gets f_{\text{beh}}(T)$

\STATE $\mathbf{z} \gets [\mathbf{v}^{\text{sem}},\ \alpha^{\text{beh}}]$
\STATE $\alpha^{\text{joint}} \gets \texttt{MLP}(\mathbf{z})$

\IF{$\alpha^{\text{joint}} \geq \theta$}
    \STATE $y^{\text{final}} \gets \texttt{Abnormal}$
\ELSE
    \STATE $y^{\text{final}} \gets \texttt{Normal}$
\ENDIF
\RETURN $y^{\text{final}}$
\end{algorithmic}
\end{algorithm}

\subsection{Inference and Decision Integration}
\label{sec:inference}
In the final stage, DMFI consolidates the outputs of the semantic and behavioral branches to produce a unified threat prediction. 
The inference process comprises three key steps: semantic scoring, behavioral scoring, and multi-feature risk fusion.

\paragraph{Semantic-Level Scoring}
For each session, we extract the semantic view $\mathcal{C} \leftarrow \{c_i \mid e_i = (t_i, a_i, o_i, d_i, c_i) \in S,\ c_i \neq \emptyset\}$, where each $c_i$ denotes a content-centric log entry (e.g., HTTP URL or email). Each entry $c_i$ is scored using the semantic LLM according to the selected DMFI strategy:
\begin{equation}
f_{\text{sem}}(c_i) =
\begin{cases}
\mathcal{M}^{\text{sem}}_{\text{mix}}(c_i), & \text{(DMFI-A)} \\
\sigma\left( \mathcal{M}^{\text{sem}}_{\text{abn}}(c_i) - \mathcal{M}^{\text{sem}}_{\text{norm}}(c_i) \right). & \text{(DMFI-B)}
\end{cases}
\end{equation}


Let $\alpha^{\text{sem}}_i = f_{\text{sem}}(c_i)$ be the anomaly score of the $i$-th semantic entry. 
In our formulation, the semantic view $\mathcal{C}$ is treated as an unordered set of content entries extracted within a session. 
Since the number of semantic entries varies across sessions, we compute descriptive statistics over the anomaly scores to produce a fixed-size and distribution-aware representation:
\begin{equation}
\mathbf{v}^{\text{sem}} = \left[
\texttt{mean}(\alpha^{\text{sem}}),
\texttt{max}(\alpha^{\text{sem}}), 
\texttt{std}(\alpha^{\text{sem}}),
\texttt{min}(\alpha^{\text{sem}})
\right].
\end{equation}

This statistical vector captures the distributional characteristics of the semantic anomalies and serves as part of the final decision input. In the edge case where $\mathcal{C} = \emptyset$ (i.e., the session contains no semantic entries), we assign a zero vector $\mathbf{v}^{\text{sem}} = [0, 0, 0, 0]$ to represent the absence of semantic information.

\paragraph{Behavioral-Level Scoring}
The behavioral log sequence is first abstracted into a natural language summary $T$ using the 4W-guided formatter $\mathcal{F}_{\text{4W}}(\cdot)$ described in Section~\ref{sec:preprocessing}. The behavioral LLM then evaluates $T$ as follows:
\begin{equation}
f_{\text{beh}}(T) =
\begin{cases}
\mathcal{M}^{\text{beh}}_{\text{mix}}(T), & \text{(DMFI-A)} \\
\sigma\left( \mathcal{M}^{\text{beh}}_{\text{abn}}(T) - \mathcal{M}^{\text{beh}}_{\text{norm}}(T) \right). & \text{(DMFI-B)}
\end{cases}
\end{equation}

Let $\alpha^{\text{beh}} = f_{\text{beh}}(T)$ denote the behavioral-level risk score. This scalar reflects the behavioral-level risk for the given session.

\paragraph{Multi-Feature Risk Fusion}
To obtain a unified threat score, we concatenate the semantic summary vector and the behavioral score into a joint feature representation:
\begin{equation}
\mathbf{z} = [\mathbf{v}^{\text{sem}},\ \alpha^{\text{beh}}].
\end{equation}

The fused vector $\mathbf{z}$ is passed through a lightweight multi-layer perceptron (MLP), which produces a final anomaly probability:
\begin{equation}
\alpha^{\text{joint}} = \texttt{MLP}(\mathbf{z}),
\end{equation}
where output $\alpha^{\text{joint}} \in [0,1]$ serves as a calibrated confidence score. A binary classification decision is derived using a threshold $\theta$:
\begin{equation}
y^{\text{final}} =
\begin{cases}
\texttt{Abnormal}, & \text{if } \alpha^{\text{joint}} \geq \theta \\
\texttt{Normal}, & \text{otherwise},
\end{cases}
\end{equation}
where $\theta = 0.5$ is used as the default threshold in the experiments.

During training, the MLP is optimized via the binary cross-entropy (BCE) loss:
\begin{equation}
\mathcal{L}_{\text{BCE}} = - \left[ y \cdot \log(\alpha^{\text{joint}}) + (1 - y) \cdot \log(1 - \alpha^{\text{joint}}) \right],
\end{equation}
where $y \in \{0,1\}$ denotes the ground-truth label for each session.

\begin{table*}[!t]
    \footnotesize
    \renewcommand{\arraystretch}{0.9}   
    \centering
    \caption{Performance comparison of different algorithms. Best results in \textbf{bold}, second-best \underline{underlined}.$\uparrow$ indicates higher values are better; $\downarrow$ indicates lower values are better.}
    \label{table:compare-method}
    \begin{tabular}{llcccc|cccc}
    \toprule
    \multirow{2}{*}{\bfseries Category} & \multirow{2}{*}{\bfseries Method} & \multicolumn{4}{c}{\bfseries CERT r4.2} & \multicolumn{4}{c}{\bfseries CERT r5.2} \\
    \cmidrule(lr){3-6} \cmidrule(lr){7-10}
    & & Prec$\uparrow$ & DR$\uparrow$ & FPR$\downarrow$ & Acc$\uparrow$ & Prec$\uparrow$ & DR$\uparrow$ & FPR$\downarrow$ & Acc$\uparrow$ \\
    \midrule

    \multirow{3}{*}{ML-Based}  &
    SVM     & 0.619 & 0.671 & 0.103 & 0.852 & 0.656 & 0.690 & 0.090 & 0.866 \\
    & IForest & 0.711 & 0.722 & 0.073 & 0.886 & 0.740 & 0.706 & 0.062 & 0.892 \\
    & XGBoost & 0.758 & 0.760 & 0.061 & 0.903 & 0.786 & 0.790 & 0.050 & 0.912 \\

    \midrule
    \multirow{4}{*}{DL-Based}  &
    Transformer & 0.782 & 0.795 & 0.056 & 0.915 & 0.814 & 0.816 & 0.047 & 0.926 \\
    & ITDBERT   & 0.832 & 0.822 & 0.042 & 0.931 & 0.850 & 0.831 & 0.037 & 0.937 \\
    & LAN       & 0.889 & 0.839 & 0.026 & 0.947 & 0.898 & 0.897 & 0.025 & 0.959 \\
    & CATE      & 0.878 & 0.862 & 0.030 & 0.948 & 0.905 & 0.929 & 0.024 & 0.966 \\

    \midrule
    \multirow{5}{*}{LLM-Based}  &
    LogGPT     & 0.910 & 0.899 & 0.022 & 0.962 & 0.895 & 0.881 & 0.026 & 0.955 \\
    & LogPrompt & 0.869 & 0.841 & 0.032 & 0.943 & 0.848 & 0.838 & 0.038 & 0.937 \\
    & Audit-LLM & \underline{0.921} &\underline{0.916} & \underline{0.015} & \underline{0.973} & \underline{0.924} & \underline{0.913} & \underline{0.015} & \underline{0.973} \\
    & ITDLM     & 0.872 & 0.839 & 0.031 & 0.943 & 0.874 & 0.860 & 0.031 & 0.947 \\
    & \bfseries DMFI-B (Ours) & \textbf{0.953} & \textbf{0.929} & \textbf{0.009} & \textbf{0.981} & \textbf{0.945} & \textbf{0.938} & \textbf{0.011} & \textbf{0.981} \\

    \bottomrule
    \end{tabular}
    \vspace{-1em}
\end{table*}

\section{Experiments}
\label{section:experiments}

\subsection{Experiments Setting}

\subsubsection{Dataset}
\label{Datasets}
In this study, we utilize the CERT Insider Threat Dataset~\cite{lindauer2020insider}, a widely adopted benchmark curated by the Software Engineering Institute (SEI) at Carnegie Mellon University. This dataset contains fine-grained user activity logs along with labeled insider threat instances and has been extensively used in prior studies on anomaly and insider threat detection.

We employ both the r4.2 and r5.2 versions of the dataset to assess the robustness and generalizability of our approach. The r4.2 dataset comprises over 32 million activity records from 1,000 users between January 2010 and May 2011, including 7,323 labeled anomalies. The r5.2 dataset contains nearly 80 million records from 2,000 users over a similar period, with 10,328 labeled anomalies.

To enable temporal behavior modeling, raw user actions are aggregated into daily session units. The dataset is partitioned into 70\% training and 30\% testing, with no user overlap to avoid data leakage. To address the severe class imbalance, we follow the undersampling strategy proposed in prior work~\cite{xiao2024unveiling,song2024audit}, limiting the number of benign sessions to 20,000 during training. We maintain a consistent session-level benign-to-anomalous ratio of approximately 8:2, balancing realism and data adequacy in line with recent LLM-based studies.

\subsubsection{Baseline Models}
\label{Comparison with Baseline Models}
We compare our method with representative baselines from three categories: (1) \textbf{ML-based models}, including SVM~\cite{le2020analyzing}, Isolation Forest~\cite{le2021anomaly}, and XGBoost~\cite{kan2023data}; (2) \textbf{DL-based models}, such as Transformer~\cite{vaswani2017attention}, ITDBERT~\cite{huang2021itdbert}, LAN~\cite{cai2024lan}, and CATE~\cite{xiao2024unveiling}; and (3) \textbf{LLM-based models}, including LogGPT~\cite{qi2023loggpt}, LogPrompt~\cite{liu2024interpretable}, Audit-LLM~\cite{song2024audit}, and ITDLM~\cite{song2025confront}. All baselines are evaluated on CERT r4.2 and r5.2 under consistent settings.

\subsubsection{Evaluation Metrics}
\label{Evaluation Metrics}
We use four standard classification metrics to evaluate detection performance: Precision (Prec), Detection Rate (DR), False Positive Rate (FPR), and Accuracy (Acc)~\cite{song2024audit,song2025confront}. Precision quantifies the proportion of correctly identified threats among predicted positives, DR (Recall) measures the coverage of true threats, FPR reflects the proportion of benign sessions incorrectly flagged, and Accuracy denotes the overall prediction correctness.

\subsubsection{Implementation Details}
The framework is implemented using LLaMA Factory~\cite{zheng2024llamafactory}, which builds on Hugging Face Transformers and the PEFT library. We use DeepSeek-R1 7B~\cite{deepseekai2025deepseekr1incentivizingreasoningcapability} as the base model and perform supervised fine-tuning (SFT) with LoRA. LoRA modules are injected into the query and value projection layers, enabling low-rank adaptation with frozen base weights.

Each training instance is formatted as an instruction-style input paired with a binary label (\texttt{Normal} or \texttt{Abnormal}), depending on the modality. Experiments are conducted on a server with four NVIDIA L20 GPUs. We use FP16 mixed-precision training and 8-bit quantization via \texttt{bitsandbytes} to improve efficiency. Training follows default configurations: LoRA rank 8, scaling factor 32, dropout rate 0.1, 3 epochs, batch size 16, AdamW optimizer with learning rate $1\times10^{-4}$, cosine scheduler, gradient accumulation (step=4), warm-up (100 steps), and gradient clipping (max norm=1.0). The final decision module is a lightweight MLP consisting of three fully connected layers. The core implementation and experimental scripts are available at \url{https://github.com/llm4cyber/dmfi-insider-threat}.

\definecolor{warmgreen}{RGB}{0, 128, 0} 
\definecolor{warmred}{RGB}{255, 69, 0}   
\newcommand{\greencheck}{\textcolor{warmgreen}{\checkmark}}
\newcommand{\redcross}{\textcolor{warmred}{\times}}

\begin{table*}[!t]
    \footnotesize
    \renewcommand{\arraystretch}{0.9}   
    \centering
    \caption{Effect of Preprocessing and Fine-Tuning Strategies. Best results in \textbf{bold}, second-best \underline{underlined}. $\uparrow$ indicates higher values are better; $\downarrow$ indicates lower values are better.}
    \label{table:ablation_modules}
    \begin{tabular}{cc|cc|cccc|cccc}
        \toprule
        \multicolumn{2}{c|}{\textbf{Preprocessing Strategy}} & \multicolumn{2}{c|}{\textbf{Fine-Tuning Strategy}} & \multicolumn{4}{c|}{\textbf{CERT r4.2}} & \multicolumn{4}{c}{\textbf{CERT r5.2}} \\
        \cmidrule(lr){1-2} \cmidrule(lr){3-4} \cmidrule(lr){5-8} \cmidrule(lr){9-12}
        \textbf{Sematic} & \textbf{Behavior} &  \textbf{DMFI-A} & \textbf{DMFI-B} & Prec $\uparrow$ & DR $\uparrow$ & FPR $\downarrow$ & Acc $\uparrow$ & Prec $\uparrow$ & DR $\uparrow$ & FPR $\downarrow$ & Acc $\uparrow$ \\
        \midrule
        $\greencheck$ & $\redcross$ &  $\greencheck$ & $\redcross$ & 0.927 & 0.911 & 0.016 & 0.973 & 0.926 & 0.916 & 0.016 & 0.971 \\
        $\greencheck$ & $\redcross$ &  $\redcross$ & $\greencheck$ & 0.936 & \underline{0.932} & 0.013 & 0.978 & \underline{0.933} & 0.931 & 0.014 & 0.976 \\
        $\redcross$ & $\greencheck$ &  $\greencheck$ & $\redcross$ & 0.902 & 0.897 & 0.021 & 0.967 & 0.893 & 0.895 & 0.022 & 0.965 \\
        $\redcross$ & $\greencheck$ &  $\redcross$ & $\greencheck$ & 0.914 & 0.902 & 0.019 & 0.970 & 0.907 & 0.897 & 0.019 & 0.968 \\
        $\greencheck$ & $\greencheck$ &  $\greencheck$ & $\redcross$ & \underline{0.942} & 0.923 & \underline{0.012} & \underline{0.979} & 0.932 & \underline{0.934} & \underline{0.013} & \underline{0.977} \\
        $\greencheck$ & $\greencheck$ &  $\redcross$ & $\greencheck$  & \textbf{0.953} & \textbf{0.929} & \textbf{0.009} & \textbf{0.981} & \textbf{0.945} & \textbf{0.938} & \textbf{0.011} & \textbf{0.981} \\
        \bottomrule
    \end{tabular}
    \vspace{-1em}
\end{table*}

\begin{table}[!t]
\footnotesize
\renewcommand{\arraystretch}{0.9}   
\centering
\caption{Effect of Semantic Score Aggregation Strategies.Best results in \textbf{bold}, second-best \underline{underlined}. $\uparrow$ indicates higher values are better; $\downarrow$ indicates lower values are better.}
\label{tab:ablation_sem}
\begin{tabular}{lcccc}
\toprule
\textbf{Method} & Prec $\uparrow$ & DR $\uparrow$ & FPR $\downarrow$  & Acc $\uparrow$ \\
\midrule
Max Only         & 0.912 & 0.926 &  0.018&  0.973\\
Mean Only        & 0.919 & \underline{0.928}  & 0.016&  0.974\\
Mean + Max       & \underline{0.932} & 0.925  &  \underline{0.013} & \underline{0.976} \\
\textbf{FullStats (Ours)} & \textbf{0.936} & \textbf{0.931} & \textbf{0.012} & \textbf{0.978} \\
\bottomrule
\end{tabular}    
\vspace{-1em}
\end{table}

\begin{figure}[!t]
    \centering
    \begin{minipage}{0.83\linewidth}
        \centering
        \subfigure[Performance on CERT r4.2]{\includegraphics[width=0.95\linewidth]{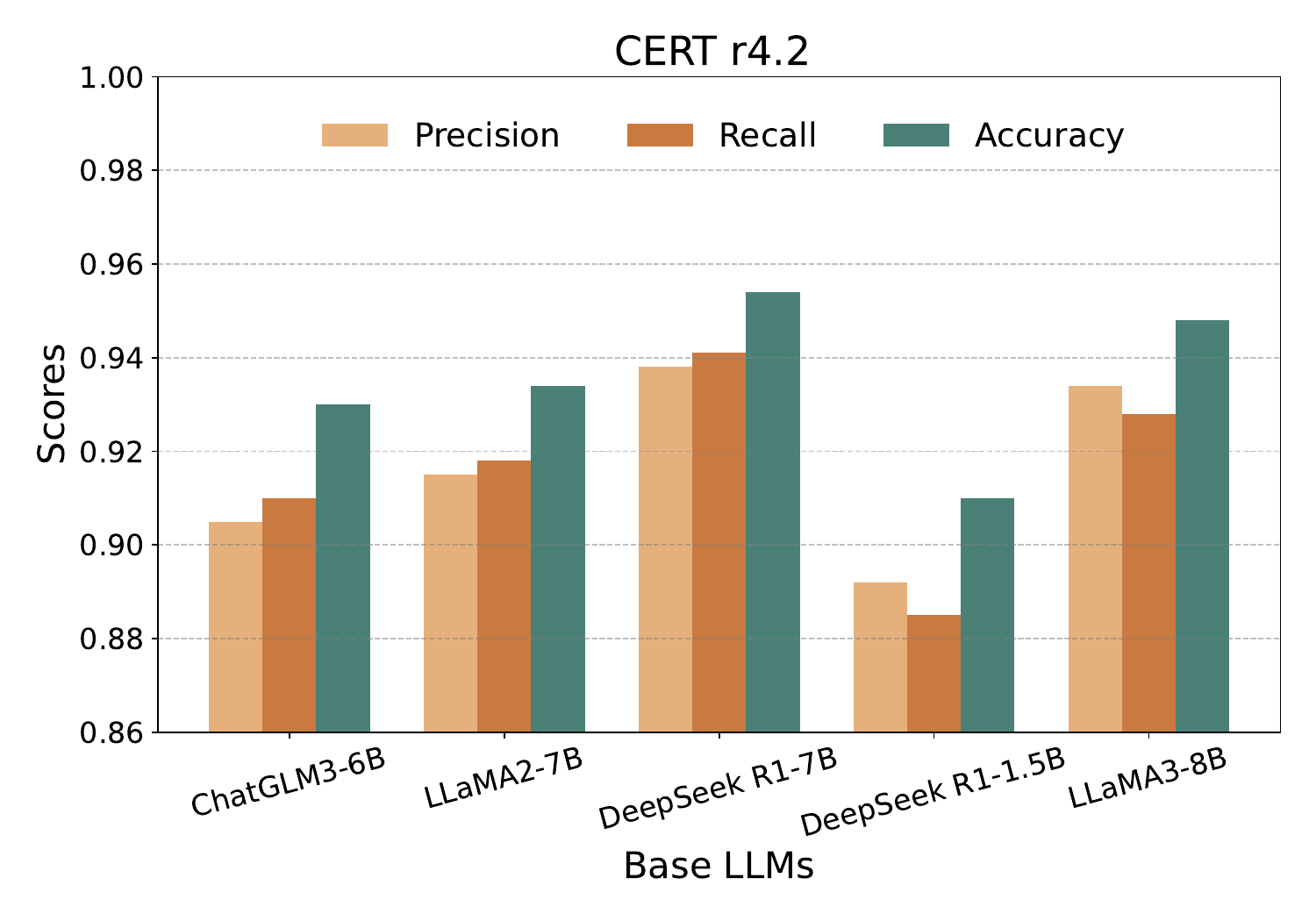}}
    \end{minipage}%
    
    \begin{minipage}{0.83\linewidth}
        \centering
        \subfigure[Performance on CERT r5.2]{\includegraphics[width=0.95\linewidth]{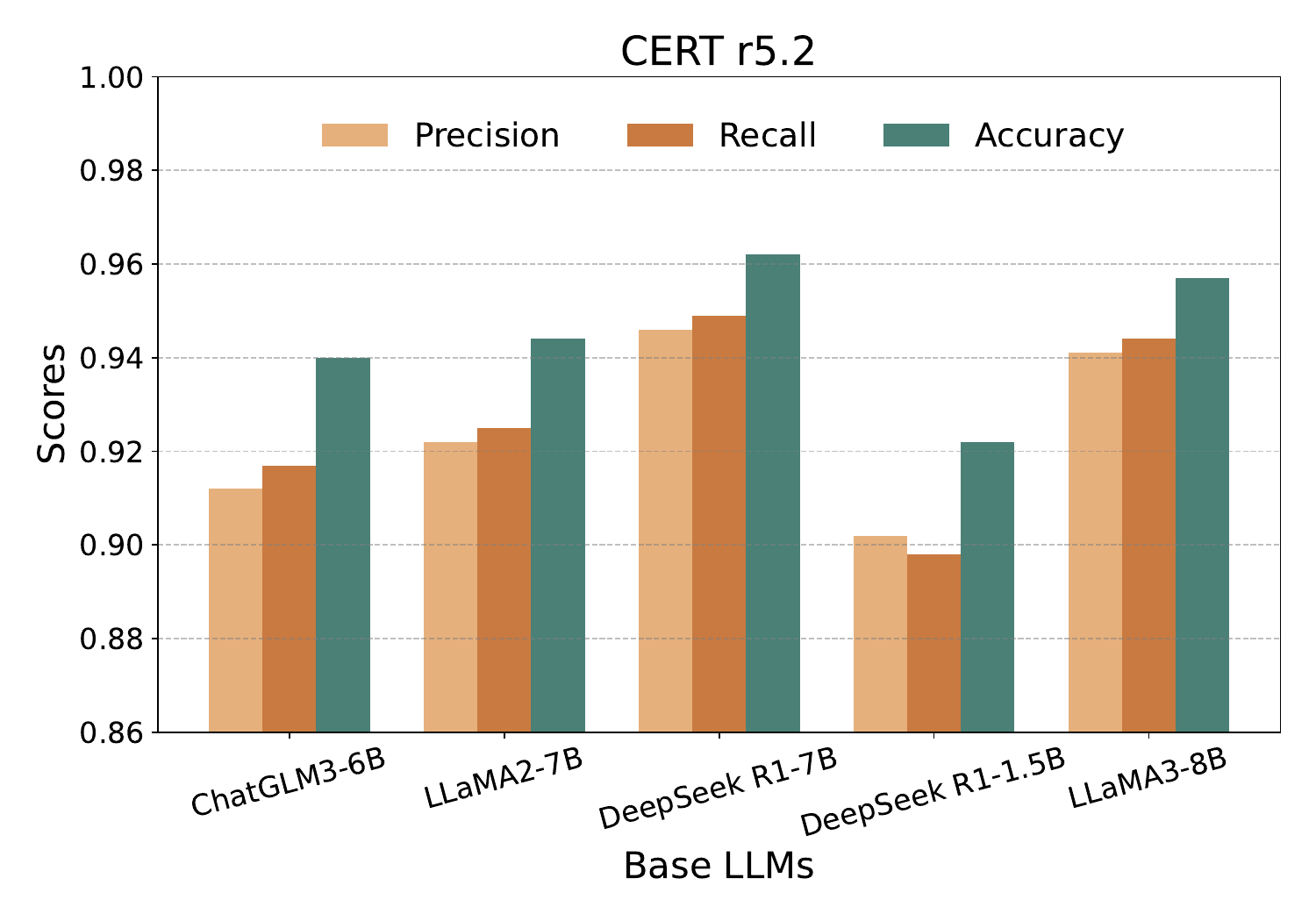}}
    \end{minipage}%
    \caption{Effect of LLM backbones on detection performance over CERT r4.2 and r5.2 datasets.}
    \label{fig:llm_comparison}
    \vspace{-1.2em}
\end{figure}

\subsection{Comparative Evaluation with Baselines}

As shown in Table~\ref{table:compare-method}, we evaluate the proposed DMFI-B framework against representative baselines from three categories: traditional ML models, deep learning (DL) architectures, and LLM-based methods. Experiments are conducted on CERT r4.2 and r5.2 datasets, using a consistent 8:2 benign-to-anomalous class ratio. Performance is assessed using Precision (Prec), Detection Rate (DR), False Positive Rate (FPR), and Accuracy (Acc).

Traditional ML models, including SVM, Isolation Forest, and XGBoost, yield moderate results with Prec ranging from 0.61 to 0.78 and FPRs up to 10.3\%. While XGBoost benefits from ensemble learning, these methods struggle under data imbalance and behavioral variability, leading to suboptimal recall and elevated false alarms.
DL-based approaches (Transformer, ITDBERT, LAN, CATE) show improved performance through sequence modeling and attention mechanisms. LAN and CATE, in particular, achieve lower FPRs (2.6\% on r4.2, 2.4\% on r5.2). However, they are less effective in handling semantic nuances or adapting to unseen behavior shifts.

DMFI-B consistently outperforms all baselines, achieving the highest Precision (0.953 / 0.945), Detection Rate (0.929 / 0.938), and lowest FPR (0.009 / 0.011) on both datasets. Compared with the second-best Audit-LLM, our method reduces FPR by over 40\% while improving recall. These results demonstrate the strength of dual-modality fine-tuning in accurately identifying insider threats with minimal false positives.

\subsection{Ablation Study}
To evaluate the contribution of each architectural component, we conduct a series of ablation experiments focusing on (i) LLM backbone selection, (ii) modality-specific preprocessing and fine-tuning modules, and (iii) semantic score aggregation strategies. These experiments are performed on both CERT r4.2 and r5.2 datasets using a consistent evaluation protocol.

\subsubsection{Effect of Backbone Language Model Choice}
We begin by investigating how the choice of language backbone influences threat detection performance. As shown in Fig.~\ref{fig:llm_comparison}, we evaluate five widely used open-source LLMs: ChatGLM3-6B, LLaMA2-7B, DeepSeek R1-7B, DeepSeek R1-1.5B, and LLaMA3-8B. These models represent a spectrum of model sizes, training alignments, and instruction-following capabilities. LLaMA3-8B and DeepSeek R1-7B consistently outperform other models across Precision, Recall, and Accuracy, indicating the benefits of strong instruction alignment and recent pretraining. Particularly, LLaMA3-8B achieves top recall on both datasets, which is critical in minimizing false negatives in insider threat scenarios.

On the other hand, the smaller DeepSeek R1-1.5B model demonstrates relatively lower performance but offers a favorable trade-off in resource-limited environments due to reduced computation cost. ChatGLM3-6B performs competitively in multilingual content but falls short in precision-heavy use cases. These results reinforce that model capacity alone is insufficient—alignment quality and instruction adaptation are essential for semantic and behavioral inference. Therefore, model selection should be based on task specificity and deployment constraints.

\begin{table}[!t]
\centering
\renewcommand{\arraystretch}{0.8}   
\setlength{\tabcolsep}{3pt} 
\scriptsize
\caption{Comparison of Raw vs. Compressed Behavior Input}
\begin{tabular}{p{1.0cm} p{4.cm} p{0.8cm} p{0.8cm}}
\toprule
\textbf{Format} & \textbf{Behavior Description} & \textbf{Token Count} & \textbf{Sentence Count} \\
\midrule
Original & During working hours, login at Self-PC.  
During working hours, at Self-PC, accessed website megaclick.com.  
During working hours, at Self-PC, accessed website linkedin.com.  
After working hours, at Self-PC, sent email from insider address to outsider address.  
After working hours, at Self-PC, opened file (.doc).  
After working hours, at Self-PC, logoff. 
& 49 & 6 \\
\midrule
4W-guided& During working hours, login at Self-PC, and then accessed multiple websites (megaclick.com, linkedin.com). After working hours, sent email (from insider address to outsider address), then opened file (.doc), and logged off. 
& 31 & 2 \\
\bottomrule
\end{tabular}
\label{tab:review-paragraph}
\vspace{-0.5em}
\end{table}

\begin{figure}[!t]
    \centering
    \subfigure[CERT r4.2]{\includegraphics[width=0.47\linewidth]{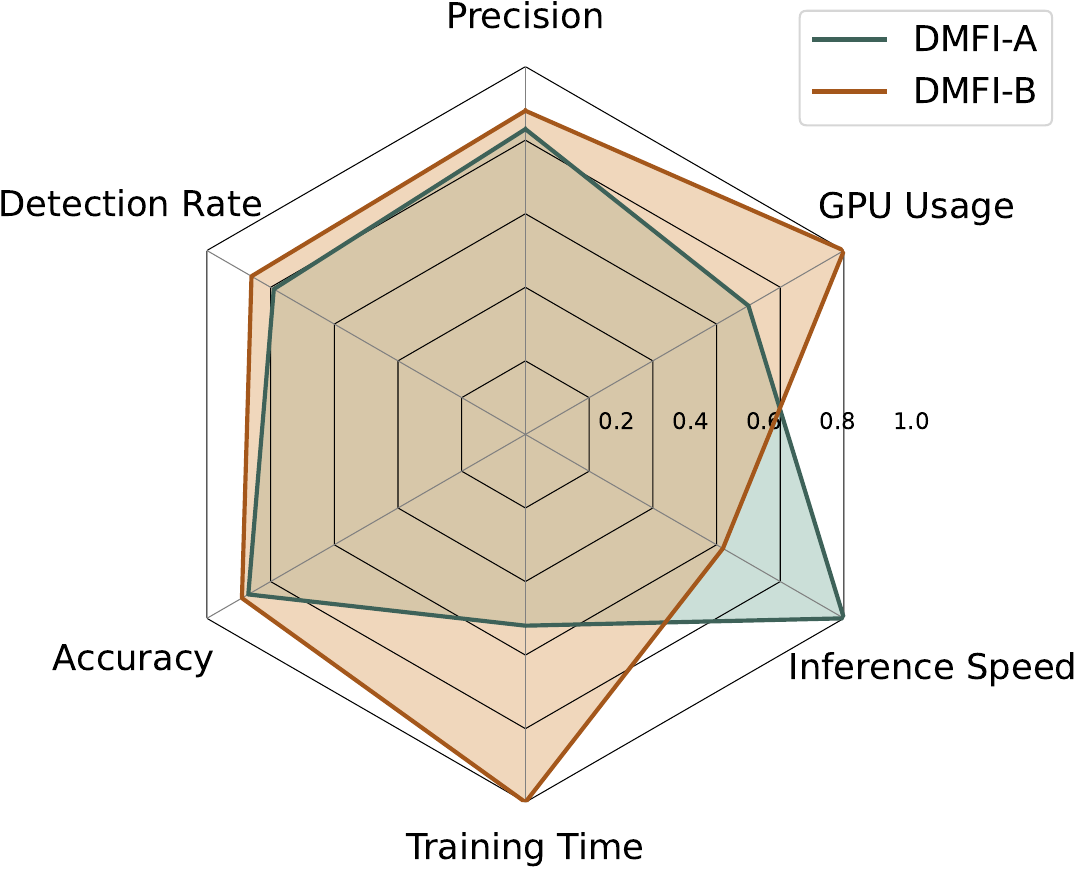}}
    \subfigure[CERT r5.2]{\includegraphics[width=0.47\linewidth]{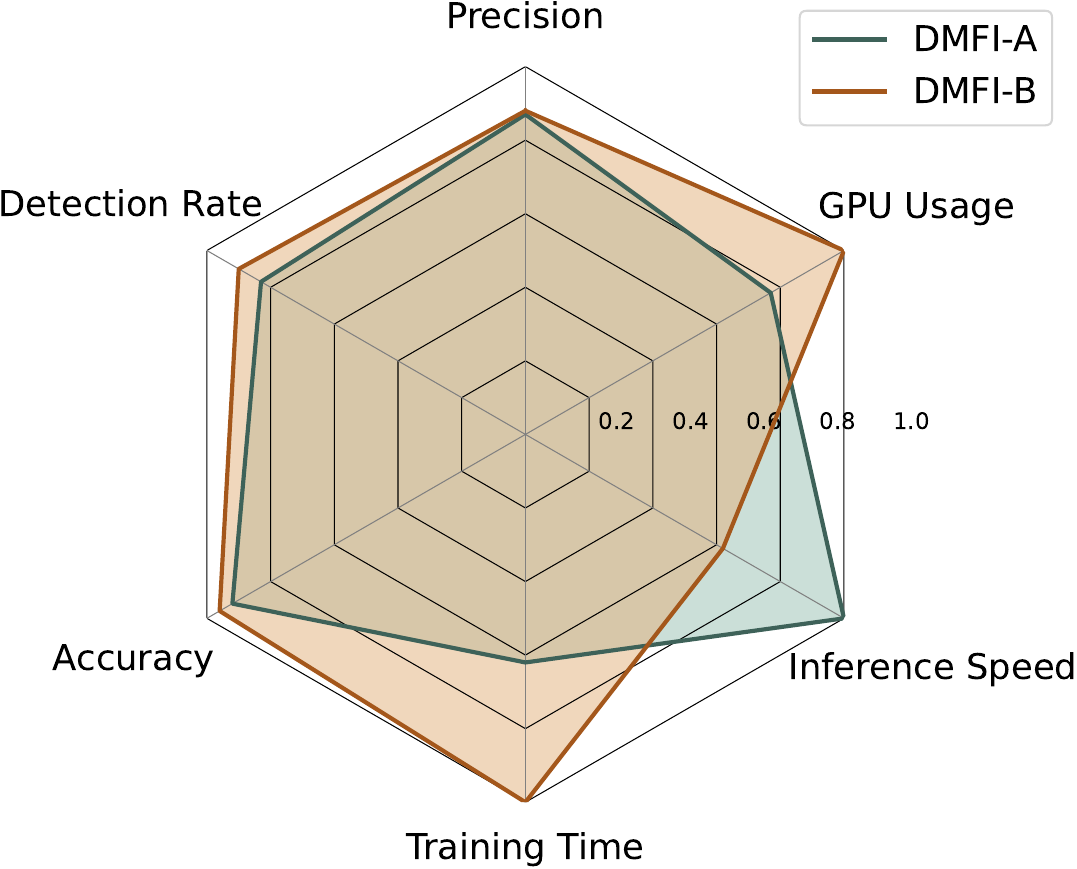}}
    \caption{Comparison of fine-tuning strategies (DMFI-A vs. DMFI-B) across performance and efficiency dimensions.}
    \label{fig:Comprehensive comparison}
    \vspace{-1em}
\end{figure}

\subsubsection{Effect of Preprocessing and Fine-Tuning Modules}
We evaluate the impact of modality-specific preprocessing and fine-tuning strategies by comparing combinations of semantic and behavioral views under two adaptation strategies: DMFI-A (unified) and DMFI-B (dual-branch). Table~\ref{table:ablation_modules} summarizes the results on CERT r4.2 and r5.2.

Incorporating both semantic and behavioral views consistently improves performance over single-view configurations. For example, removing either view leads to a 2–4\% drop in Accuracy and Detection Rate. This confirms the complementary nature of content-level semantics and behavior-level structure.

Additionally, DMFI-B outperforms DMFI-A across all configurations. Its discriminative dual-branch structure improves anomaly localization by separately modeling normal and abnormal data. When both views are present, DMFI-B improves Accuracy by approximately 0.4\% and reduces FPR by 0.002–0.003 compared to DMFI-A. These results hold consistently across both datasets, validating the effectiveness of dual-view representation and fine-grained adaptation.

\subsubsection{Effect of Semantic Score Aggregation Strategies}
Lastly, we analyze the semantic score aggregation strategy, which transforms variable-length semantic predictions ${\alpha^{\text{sem}}_i}$ into a fixed-dimensional representation for final classification. Table~\ref{tab:ablation_sem} presents the results under four methods: Max Only, Mean Only, Mean + Max, and FullStats (ours). Max pooling captures the most salient anomaly but ignores the overall distribution, making it vulnerable to local noise. In contrast, using the mean alone improves robustness but can mask sharp deviations, reducing sensitivity to localized threat spikes.

By integrating both mean and max, the model gains a more balanced view of anomaly magnitude and scope. Our proposed FullStats method further incorporates standard deviation and minimum values, encoding variance and suppression patterns. This richer representation consistently achieves the best trade-offs in all metrics, with improvements of up to 0.008 in accuracy and 0.011 in FPR over the second-best baseline. The results demonstrate that multi-statistic vectors capture more nuanced risk profiles and support better-calibrated predictions for downstream decision-making.

\subsection{Efficiency-Effectiveness Trade-off Analysis}
To assess the trade-offs between detection effectiveness and computational cost, we conduct two targeted evaluations: (1) the impact of compressing behavior sequences, and (2) the efficiency-performance balance between fine-tuning strategies.

\subsubsection{Impact of Behavior Compression}
Given that LLMs are sensitive to input length and context window size, we examine the benefits of compressing verbose behavior logs into concise narrative representations. Table~\ref{tab:review-paragraph} illustrates a representative session from user \texttt{AAM0658} on \textbf{January 15, 2010}, spanning actions during and after working hours.

The \textbf{Original} format describes each action independently, resulting in 49 tokens and 6 sentences. In contrast, the \textbf{4W-guided} format applies our abstraction function to merge semantically related events, reducing the input to 31 tokens and 2 sentences—a 36.7\% reduction in token count. Importantly, critical indicators such as off-hour file access and external communication are preserved. This abstraction not only enhances inference efficiency but also improves interpretability, which is essential for real-world deployments constrained by context length or throughput.

\subsubsection{Resource Efficiency of Fine-Tuning Strategies}
We further analyze the efficiency-performance trade-off between the unified fine-tuning approach (DMFI-A) and the dual-branch strategy (DMFI-B). While DMFI-B consistently yields superior performance across Precision, Detection Rate, and Accuracy (as shown in Fig.~\ref{fig:Comprehensive comparison}), this improvement incurs additional training and inference costs.
DMFI-B maintains separate pathways for normal and abnormal sessions, enabling better specialization through LoRA-based adaptation. However, this dual-branch design results in increased GPU memory usage and longer runtime. Conversely, DMFI-A offers a lighter architecture with faster inference and reduced training time, making it more suitable for resource-constrained deployments.

These results underscore the flexibility of DMFI in balancing detection fidelity with runtime efficiency, and provide practical guidance for model selection based on infrastructure and latency constraints.

\begin{table}[!t]
\centering
\renewcommand{\arraystretch}{1.3}   
\setlength{\tabcolsep}{4pt}
\caption{Case Study of Semantic and Behavioral Integration}
\label{tab:casestudy}
\scriptsize 
\begin{tabular}{|p{1.2cm}|p{7.1cm}|}

\hline
\rowcolor{black!10} 
\textbf{Log} & 
[01:34, PC-9923] Logon \newline
[04:04, PC-9923] access http://amazon.com/ \quad  \newline \textit{content:} provisions and explorer should... \newline
[04:35, PC-9923] access http://amazon.com/ \quad \newline \textit{content:}design through draws private... \newline
[06:18, PC-9923] Connect device \newline
[06:26, PC-9923] access http://wikileaks.org/ \quad \newline \textit{content:}spy bait bait distort evade... \newline
[06:26, PC-9923] Disconnect device \newline
[06:27, PC-9923] access http://amazon.com/ \quad \newline \textit{content:}native 1967 first 1867 brought... \newline
[06:28, PC-9923] Logoff \\
\hline

\rowcolor{blue!5}
\textbf{Semantic Branch} & 
$C_1$: provisions and explorer should...$\rightarrow$  $\alpha^{\text{sem}}_1 = \textbf{0.13, Normal} $ \newline
$C_2$: design through draws private... $\rightarrow$  $\alpha^{\text{sem}}_2 = \textbf{0.07, Normal}$  \newline
$C_3$: spy bait bait distort evade... $\rightarrow$  $\alpha^{\text{sem}}_3 = \textbf{\textcolor{red}{0.83, Abnormal}} $ \newline
$C_4$: native 1967 first 1867 brought... $\rightarrow$  $\alpha^{\text{sem}}_4 = \textbf{0.21, Normal} $ \\
\hline

\rowcolor{green!5}
\textbf{Behavioral Branch} & 
$T$: After working hours, login at Self-PC, then accessed multiple websites (amazon.com, wikileaks.org), connected and disconnected an external device, revisited amazon.com, and logged off.  $\rightarrow$ 
$\alpha^{beh} = \textbf{\textcolor{red}{0.93, Abnormal}}$ \\
\hline

\rowcolor{yellow!10}
\textbf{Fusion} & 
$\mathbf{v}^{sem} = [0.31, 0.83, 0.30, 0.07]$, \quad $\alpha^{beh} = 0.93$ \newline
$\mathbf{z} = [\mathbf{v}^{sem}, \alpha^{beh}] = [0.31, 0.83, 0.30, 0.07, 0.93]$ \\
\hline

\rowcolor{red!10}
\textbf{Result} & 
$y^{final} = \textbf{0.91}$ \quad Decision: \textbf{\textcolor{red}{Abnormal}} \\
\hline
\end{tabular}
\vspace{-1.5em}
\end{table}

\subsection{Case Study}
As shown in Table~\ref{tab:casestudy}, 
this case study demonstrates the workflow of our proposed DMFI framework. 
Starting from raw activity logs, the semantic branch evaluates content-level anomalies, 
while the behavioral branch summarizes sequential actions into interpretable natural language and produces a session-level score. 
Although most semantic entries appear normal, the access to \texttt{wikileaks.org} yields a high anomaly score ($\alpha^{sem}_3 = 0.83$), 
and the behavioral branch further identifies abnormal behavioral patterns ($\alpha^{beh} = 0.93$). 
By fusing both perspectives, the joint prediction reaches $y^{final} = 0.91$, 
leading to the final decision of \textbf{Abnormal}. 
This example illustrates how our framework integrates semantic reasoning and behavioral summarization into a unified, interpretable decision process.

\section{Conclusion and Future Work}
\label{section:conclusion}
In this paper, we proposed DMFI, a dual-modality framework for insider threat detection that combines prompt-based semantic reasoning and behavior-aware fine-tuning within a unified architecture. Our design enables DMFI to model both content-level semantics and temporal behavioral dependencies effectively, addressing the challenges of intent ambiguity and long-range activity patterns.
We introduced two complementary fine-tuning strategies: DMFI-A for efficient deployment and DMFI-B for enhanced anomaly sensitivity under class imbalance. Additionally, we presented a multi-statistic fusion mechanism to integrate heterogeneous inference signals across user sessions. Evaluations on CERT r4.2 and r5.2 datasets confirm that DMFI outperforms both traditional models and recent LLM-based detectors in precision, generalization, and computational efficiency.

In future work, we plan to (1) extend DMFI to cross-domain environments such as cloud audit logs and IoT telemetry, (2) explore human-in-the-loop strategies for interactive prompt optimization, and (3) incorporate privacy-preserving training objectives (e.g., federated adaptation or differential privacy) for real-world deployment in sensitive domains.


\begin{thebibliography}{00}

\bibitem{fei2025laaeb}
K. Fei, J. Zhou, Y. Zhou, X. Gu, H. Fan, B. Li, W. Wang, and Y. Chen, 
``LaAeb: A comprehensive log-text analysis based approach for insider threat detection,'' 
\textit{Computers \& Security}, vol. 148, p. 104126, 2025.

\bibitem{inayat2024insider}
U. Inayat, M. Farzan, S. Mahmood, M. F. Zia, S. Hussain, and F. Pallonetto, 
``Insider threat mitigation: Systematic literature review,'' 
\textit{Ain Shams Engineering Journal}, p. 103068, 2024.

\bibitem{liu2018detecting}
L. Liu, O. De Vel, Q.-L. Han, J. Zhang, and Y. Xiang, 
``Detecting and preventing cyber insider threats: A survey,'' 
\textit{IEEE Communications Surveys \& Tutorials}, vol. 20, no. 2, pp. 1397--1417, 2018.

\bibitem{le2021anomaly}
D. C. Le and N. Zincir-Heywood, 
``Anomaly detection for insider threats using unsupervised ensembles,'' 
\textit{IEEE Transactions on Network and Service Management}, vol. 18, no. 2, pp. 1152--1164, 2021.

\bibitem{al2020review}
M. N. Al-Mhiqani, R. Ahmad, Z. Z. Abidin, W. Yassin, A. Hassan, K. H. Abdulkareem, N. S. Ali, and Z. Yunos, 
``A review of insider threat detection: Classification, machine learning techniques, datasets, open challenges, and recommendations,'' 
\textit{Applied Sciences}, vol. 10, no. 15, p. 5208, 2020.

\bibitem{le2020analyzing}
D. C. Le, N. Zincir-Heywood, and M. I. Heywood, 
``Analyzing data granularity levels for insider threat detection using machine learning,'' 
\textit{IEEE Transactions on Network and Service Management}, vol. 17, no. 1, pp. 30--44, 2020.

\bibitem{kan2023data}
X. Kan, Y. Fan, J. Zheng, C.-h. Chi, W. Song, and A. Kudreyko, 
``Data adjusting strategy and optimized XGBoost algorithm for novel insider threat detection model,'' 
\textit{Journal of the Franklin Institute}, vol. 360, no. 16, pp. 11414--11443, 2023.

\bibitem{al2024comparative}
T. Al-Shehari, M. Kadrie, M. N. Al-Mhiqani, T. Alfakih, H. Alsalman, M. Uddin, S. S. Ullah, and A. Dandoush, 
``Comparative evaluation of data imbalance addressing techniques for CNN-based insider threat detection,'' 
\textit{Scientific Reports}, vol. 14, no. 1, p. 24715, 2024.

\bibitem{xiao2024sentinel}
F. Xiao, S. Chen, S. Chen, Y. Ma, H. He, and J. Yang, 
``SENTINEL: Insider threat detection based on multi-timescale user behavior interaction graph learning,'' 
\textit{IEEE Transactions on Network Science and Engineering}, 2024.

\bibitem{yuan2021deep}
S. Yuan and X. Wu, 
``Deep learning for insider threat detection: Review, challenges and opportunities,'' 
\textit{Computers \& Security}, vol. 104, p. 102221, 2021.

\bibitem{chen2024survey}
Y. Chen, M. Cui, D. Wang, Y. Cao, P. Yang, B. Jiang, Z. Lu, and B. Liu, 
``A survey of large language models for cyber threat detection,'' 
\textit{Computers \& Security}, p. 104016, 2024.

\bibitem{qi2023loggpt}
J. Qi, S. Huang, Z. Luan, S. Yang, C. Fung, H. Yang, D. Qian, J. Shang, Z. Xiao, and Z. Wu, 
``LogGPT: Exploring ChatGPT for log-based anomaly detection,'' 
in \textit{Proc. IEEE Int. Conf. High Performance Computing \& Communications, Data Science \& Systems, Smart City \& Dependability in Sensor, Cloud \& Big Data Systems \& Application (HPCC/DSS/SmartCity/DependSys)}, 2023, pp. 273--280.

\bibitem{li2025redchronos}
C. Li, Z. Zhu, J. He, and X. Zhang, 
``RedChronos: A large language model-based log analysis system for insider threat detection in enterprises,'' 
\textit{arXiv preprint arXiv:2503.02702}, 2025.

\bibitem{song2024audit}
C. Song, L. Ma, J. Zheng, J. Liao, H. Kuang, and L. Yang, 
``Audit-LLM: Multi-agent collaboration for log-based insider threat detection,'' 
\textit{arXiv preprint arXiv:2408.08902}, 2024.

\bibitem{song2025confront}
S. Song, Y. Zhang, and N. Gao, 
``Confront insider threat: Precise anomaly detection in behavior logs based on LLM fine-tuning,'' 
in \textit{Proc. 31st Int. Conf. Computational Linguistics (COLING)}, 2025, pp. 8589--8601.

\bibitem{karlsen2024benchmarking}
E. Karlsen, X. Luo, N. Zincir-Heywood, and M. Heywood, 
``Benchmarking large language models for log analysis, security, and interpretation,'' 
\textit{Journal of Network and Systems Management}, vol. 32, no. 3, p. 59, 2024.

\bibitem{wang2024feditd}
Z. Q. Wang, H. Wang, and A. El Saddik, 
``FedITD: A federated parameter-efficient tuning with pre-trained large language models and transfer learning framework for insider threat detection,'' 
\textit{IEEE Access}, 2024.

\bibitem{homoliak2019insight}
I. Homoliak, F. Toffalini, J. Guarnizo, Y. Elovici, and M. Ochoa, 
``Insight into insiders and IT: A survey of insider threat taxonomies, analysis, modeling, and countermeasures,'' 
\textit{ACM Computing Surveys (CSUR)}, vol. 52, no. 2, pp. 1--40, 2019.

\bibitem{villarreal2021hunting}
M. Villarreal-Vasquez, G. Modelo-Howard, S. Dube, and B. Bhargava, 
``Hunting for insider threats using LSTM-based anomaly detection,'' 
\textit{IEEE Transactions on Dependable and Secure Computing}, vol. 20, no. 1, pp. 451--462, 2021.

\bibitem{zhu2024auth}
X. Zhu, J. Dong, J. Qi, Z. Zhou, Z. Dong, Y. Sun, and M. Wang, 
``AUTH: An adversarial autoencoder based unsupervised insider threat detection scheme for multisource logs,'' 
\textit{IEEE Transactions on Industrial Informatics}, 2024.

\bibitem{clairoux2024use}
V. Clairoux-Trepanier, I.-M. Beauchamp, E. Ruellan, M. Paquet-Clouston, S.-O. Paquette, and E. Clay, 
``The use of large language models (LLM) for cyber threat intelligence (CTI) in cybercrime forums,'' 
\textit{arXiv preprint arXiv:2408.03354}, 2024.

\bibitem{gelman2025scalable}
H. Gelman and J. D. Hastings, 
``Scalable and ethical insider threat detection through data synthesis and analysis by LLMs,'' 
in \textit{Proc. 13th Int. Symp. on Digital Forensics and Security (ISDFS)}, IEEE, 2025, pp. 1--6.

\bibitem{narvala2023identifying}
H. Narvala, G. McDonald, and I. Ounis, 
``Identifying chronological and coherent information threads using 5W1H questions and temporal relationships,'' 
\textit{Information Processing \& Management}, vol. 60, no. 3, p. 103274, 2023.

\bibitem{mao2025survey}
Y. Mao, Y. Ge, Y. Fan, W. Xu, Y. Mi, Z. Hu, and Y. Gao, 
``A survey on LoRA of large language models,'' 
\textit{Frontiers of Computer Science}, vol. 19, no. 7, p. 197605, 2025.

\bibitem{lindauer2020insider}
B. Lindauer, 
``Insider threat test dataset,'' 
\textit{doi:10.1184/R1/12841247.v1}, 2020. [Online]. Available: https://doi.org/10.1184/R1/12841247.v1.

\bibitem{xiao2024unveiling}
H. Xiao, Y. Zhu, B. Zhang, Z. Lu, D. Du, and Y. Liu, 
``Unveiling shadows: A comprehensive framework for insider threat detection based on statistical and sequential analysis,'' 
\textit{Computers \& Security}, vol. 138, p. 103665, 2024.

\bibitem{vaswani2017attention}
A. Vaswani, N. Shazeer, N. Parmar, J. Uszkoreit, L. Jones, A. N. Gomez, Ł. Kaiser, and I. Polosukhin, 
``Attention is all you need,'' 
in \textit{Advances in Neural Information Processing Systems (NeurIPS)}, 2017.

\bibitem{huang2021itdbert}
W. Huang, H. Zhu, C. Li, Q. Lv, Y. Wang, and H. Yang, 
``ITDBERT: Temporal-semantic representation for insider threat detection,'' 
in \textit{Proc. 2021 IEEE Symp. Computers and Communications (ISCC)}, 2021, pp. 1--7.

\bibitem{cai2024lan}
X. Cai, Y. Wang, S. Xu, H. Li, Y. Zhang, Z. Liu, and X. Yuan, 
``LAN: Learning adaptive neighbors for real-time insider threat detection,'' 
\textit{IEEE Transactions on Information Forensics and Security}, 2024.

\bibitem{liu2024interpretable}
Y. Liu, S. Tao, W. Meng, J. Wang, W. Ma, Y. Chen, Y. Zhao, H. Yang, and Y. Jiang, 
``Interpretable online log analysis using large language models with prompt strategies,'' 
in \textit{Proc. 32nd IEEE/ACM Int. Conf. Program Comprehension (ICPC)}, 2024, pp. 35--46.


\bibitem{zheng2024llamafactory}
Y. Zheng, R. Zhang, J. Zhang, Y. Ye, Z. Luo, Z. Feng, and Y. Ma, 
``LlamaFactory: Unified efficient fine-tuning of 100+ language models,'' 
\textit{arXiv preprint arXiv:2403.13372}, 2024.


\bibitem{deepseekai2025deepseekr1incentivizingreasoningcapability}
D. Guo, D. Yang, H. Zhang, J. Song, R. Zhang, R. Xu, Q. Zhu, S. Ma, P. Wang, X. Bi, \textit{et al.}, 
``DeepSeek-R1: Incentivizing reasoning capability in LLMs via reinforcement learning,'' 
\textit{arXiv preprint arXiv:2501.12948}, 2025.



\end{thebibliography}
\end{document}